\documentclass[preprint,aps,nofootinbib,superscriptaddress,showpacs,showkeys]{revtex4}
\usepackage{epsfig}
\usepackage{amssymb}
\usepackage{amsmath}
\usepackage{bm}
\usepackage{graphicx}
\usepackage{hyperref}

\begin{document}

\title{Percentage depth dose distributions in inhomogeneous phantoms 
with lung and bone equivalent media for small fields of CyberKnife}

\author{Chung Il Lee} 
\affiliation{Department of Radiation Oncology, The Catholic University of Korea, 
Bucheon St. Mary's Hospital, Bucheon, Korea}

\author{Jae Won Shin}
\affiliation{Department of Physics, Sungkyunkwan University,
Suwon, Korea}

\author{Sei-Chul Yoon}
\affiliation{Department of Radiation Oncology, The Catholic University of Korea, 
Bucheon St. Mary's Hospital, Bucheon, Korea}

\author{Tae Suk Suh}
\affiliation{Department of Biomedical Engineering, the Catholic University, Seoul, Korea}

\author{Seung-Woo Hong}
\affiliation{Department of Physics, Sungkyunkwan University,
Suwon, Korea}
\affiliation{Department of Energy Science, 
Sungkyunkwan University, Suwon, Korea}

\author{Kyung Joo Min}
\affiliation{Department of Energy Science, 
Sungkyunkwan University, Suwon, Korea}

\author{Sang Deok Lee}
\affiliation{Department of Physics, Sungkyunkwan University,
Suwon, Korea}
\affiliation{Department of Optometry, Gimcheon University, Gimcheon, Korea}

\author{Su Mi Chung}
\affiliation{Department of Radiation Oncology, The Catholic University of Korea, 
Yeouido St. Mary's Hospital, Seoul, Korea}

\author{Jae-Yong Jung}
\affiliation{Department of Radiation Oncology, Sanggye Paik Hospital, 
Inje University, Seoul, Korea}

\date{31 December 2013 }

\begin{abstract}

The percentage depth dose distributions in inhomogeneous phantoms 
with lung and bone equivalent media are studied. 
For lung equivalent media a Balsa wood is used, 
and for a soft bone equivalent media a compound material with epoxy resin, 
hardener and calcium carbonate is used. 
Polystyrene slabs put together with these materials are used as an inhomogeneous phantom. 
Dose measurements are performed with Gafchromic EBT film by using photon beams from 
6MV CyberKnife at the Seoul Uridul Hospital. 
The cone sizes of the photon beams are varied 
from 5, 10 to 30 mm. 
As a simulation tool GEANT4 Monte Carlo code v9.4.p02 is used. 
When the Balsa wood is inserted in the phantom, the dose measured with 
EBT film is found to be significantly different from the dose without the EBT film in 
and beyond the Balsa wood region, particularly for small field sizes. 
On the other hand, when the soft bone equivalent material is inserted in the phantom, 
discrepancy between the dose measured with EBT film and the dose without 
EBT film can be seen only in the region of bone equivalent material. 
GEANT4 simulations are done with and without the EBT film 
to compare the simulation results with measurements. 
We find that the simulations including EBT film agree with the measurements 
for all the cases within an error of 2.2\%. 
Also, we find the "doses to phantom" 
without the EBT film differ from the "doses to film" up to 29\%, 
which shows that for accurate dose estimations for inhomogeneous phantoms with 
EBT film the presence of the EBT film needs to be 
taken into account properly in particular for small fields.

\end{abstract}



\maketitle

\section{INTRODUCTION}

Stereotactic radiosurgery (SRS) \cite{r_1} uses small field sizes to deliver 
the maximum dose to the target and the minimum dose to its surrounding normal tissues. 
To facilitate the use of these small field sizes, specialized equipments such as 
CyberKnife \cite{r_2}, Gamma knife \cite{r_3}, 
multileaf collimating systems \cite{r_4, r_5} and etc. are needed. 
A higher accuracy is needed compared to conventional radiotherapy to deliver photon beams 
with the uncertainty of $\pm$ 1 mm at the target \cite{r_6}. 
However, measurements of the doses for small fields, often used in SRS, are difficult due to 
the absence of electronic equilibrium in radiation fields of dimensions 
smaller than the maximum range of secondary electrons. 
Such difficulties associated with small beam fields bring about uncertainties 
in the area of clinical dosimetry.

Different dosimeters are used for small field radiations: 
ionization chambers, diamond detectors, silicon diodes, radiochromic films and etc. 
Ionization chambers have been a standard of the dose measurement 
and calibrations for general radiation treatment. 
However, ionization chambers are larger than 
the small beam field sizes used in SRS. 
Silicon diodes are often selected for stereotactic radiotherapy (SRT) and SRS 
because of its small sensitive volumes. 
However, it is energy-, dose rate- and direction-dependent \cite{r_7, r_8}. 
Silicons which have a higher atomic number than tissues are sensitive to low energy photons 
and thus tend to overestimate the dose at low energies. 
Diamond detectors which have been developed for photon dosimetry 
for small fields are tissue equivalent. 
Though its spatial resolution is good, 
it is very expensive and dose-rate dependent \cite{r_9, r_10}. 
Gafchromic EBT dosimetry films are near tissue-equivalent 
and independent of photon energy from keV to MeV ranges. 
Because it has a high spatial resolution and does not require film development, 
it is used for quality assurance of Intensity-Modulated Radiation Therapy (IMRT) 
and photon detection for radiosurgery of small field sizes \cite{r_11, r_12},
though the sensitivity and accuracy of measurement could differ 
by the choice of densitometry system and calibration \cite{r_13, r_14, r_15, r_16}. 
The EBT films, diode detectors and ion chambers produce 
nearly identical values of percentage depth dose (PDD) 
for field sizes larger than 10 mm in CyberKnife. 
However, for field sizes of 5 mm and 7.5 mm, 
these detectors do not produce identical PDDs \cite{r_12}. 
Detailed comparisons of these detectors can be found in Table 1 of Ref. \cite{r_12}.

When SRS/SRT was initially implemented to the treatment of brain, 
inhomogeneity of the brain was not seriously considered. 
As the use of CyberKnife expands, however, 
more accurate dose estimation is needed 
for tissues with inhomogeneous region, such as air cavity, lung and bones. 
Agreement between the treatment plan 
and the measured dose values for small field radiations 
in inhomogeneous region in spine and lung tissues 
is a major factor that affects treatment accuracy. 
Techniques to correct for inhomogeneity 
have been developed with increasing computing power 
and improved understanding of density-variable regions \cite{r_17}. 
Accurate dose measurement is particularly important in the dosimetry of small fields 
in the presence of low density inhomogeneities. 
When the photon beams pass through the air cavity, 
which is the lowest density area in the human body, 
the dose drops in the air cavity, 
whereas it gets built-up in the phantom, 
and thus a dose enhancement may be developed 
in the distal side of cavity \cite{r_17, r_18, r_19, r_20}.

The characteristics of the inhomogeneous conditions 
were studied with the radiochromic film 
type MD-55 (Gafchromic type MD-55 Cat no 37-041 Lot no J1548MD55) 
by Paelinck {\em et al}. \cite{r_18, r_21}. 
They performed the dose measurement with films and Monte Carlo simulations for 
inhomogeneous phantoms involving the air cavity or lung equivalent media 
with the photon beams provided by 6MV linear accelerator. 
In their simulations, they modeled the phantoms 
and the film by using BEAMnrc/EGSnrc system 
and then compared the simulation results with measurements, 
obtaining consistency between the film measurements and 
the calculations with including the film. 
The dose calculations were performed in two different methods. 
First, the simulations were performed with a film detector, 
and thus the doses were obtained from the film. 
We refer to these values of dose as the "dose to film (DTF)". 
Second, the simulations were done with phantom geometry without including the film. 
In this case, the absorbed doses were calculated 
in the phantom material itself without 
taking into account the presence of the film. 
So, we call these values of dose the "dose to phantom (DTP)". 
For the case when the air cavity was inserted in the phantom, 
they found a difference between the DTF and the DTP \cite{r_18, r_21}. 
They showed that the differences decrease 
as the field size and the density of the medium increase. 
Also they found that these differences disappear by offsetting 
the film sideway by a few centimeters.

A study of PDD for inhomogeneous phantoms 
with a Balsa wood as a low density material 
or a cortical bone as a high density material 
was done by Wilcox and Daskalov \cite{r_22}, 
who compared the EBT film measurements of PDD (DTF) with the calculations of DTP 
for inhomogeneous conditions by using Cyberknife systems. 
In that work, they did not include the presence of the EBT film in the simulation 
since the film thickness was smaller than the pixel size of the CT image. 
Thus they compared the calculated DTP with the measured DTF. 
In the region of 
low density materials such as a Balsa wood, 
calculations ignoring the film material 
did not make a noticeable difference. 
Beyond the Balsa wood region, however, 
the calculated DTP was found to be larger than the measured DTF \cite{r_22}. 
Similar behavior can be observed in the works of Paelinck {\em et al}. \cite{r_18, r_21}. 
For inhomogeneous phantoms with high density materials, discrepancies between 
the DTF and the DTP were observed in the cortical bone region. 
The ratios of the DTF to the DTP in cortical bone regions were consistent 
with the correction factors suggested by Siebers {\em et al}. \cite{r_23}.

The discrepancy between the DTP 
and the DTF increases as the field size decreases. 
Thus, if the beam field size becomes smaller than 
previously considered \cite{r_18, r_21, r_22}, 
it can make more difference between the DTP 
and the DTF in inhomogeneous conditions. 
At present, 5 mm is the smallest 
cone size of the CyberKnife systems. 
Previously, due to the rare use of 5 mm cone size 
in clinical practice, 
this small cone size was not considered seriously. 
However, as an extension of the previous studies 
of the inhomogeneous conditions \cite{r_18, r_21, r_22}, 
we have considered in this work low and high density materials 
for inhomogeneous phantoms 
with 5, 10 and 30 mm cone sizes of the photon beams 
produced by 6MV CyberKnife and performed 
both simulations and measurements of PDD.

For the calculation of PDD, we used GEANT4 v9.4.p02 \cite{r_24}, 
which can simulate particle transport 
in matter and calculate the dose. Carrier {\em et al}. \cite{r_25} 
validated the GEANT4 v5.2 for electron 
and photon transportation in homogeneous and multilayer phantoms. 
They showed that the doses obtained by GEANT4 results 
are comparable to those obtained by using 
other commonly used codes, such as MCNP, EGSnrc and EGS4. 
Poon {\em et al}. \cite{r_26, r_27} 
performed a test of consistency of GEANT4 v6.2.p01 against EGS4 and the existing data. 
Elles {\em et al}. \cite{r_28} found that an improvement has been made in the version 8.3 of GEANT4 
by employing the same conditions as used by Poon {\em et al}. \cite{r_26, r_27}. 
Also, we applied GEANT4 v9.1 to a $^{60}$Co therapy unit 
and calculated PDD distributions, 
peak scatter factors and tissue air ratios \cite{r_29}. 
We obtained consistency 
between the calculations and the published data (BJR suppl.25) \cite{r_30} 
with errors less than 1\%. 
In addition, we have simulated photon beams produced by 6MV CyberKnife system 
at the Seoul Uridul Hospital with GEANT4 v9.2.p01 \cite{r_31}, 
where we found that the calculated values of PDD distributions, output factors 
and off axis ratios agree well with the experimental values 
measured with silicon diode detectors within 2\% errors.

In this work, we consider an inhomogeneous phantom with spine or 
lung equivalent materials and compare 
the calculated absorbed doses with the measured ones. 
Measurements have been done with Gafchromic EBT2 films. 
To see the effect of the presence of 
the EBT2 film on the PDD, we have performed the simulations with 
and without the EBT2 films. 
Also, we compared the measured values of PDD 
with the calculated values using Multiplan 
which is the software program used in CyberKnife treatment systems.

\section{METHODS}

\subsection{Experiment}

\begin{figure}[tbp]
\epsfig{file=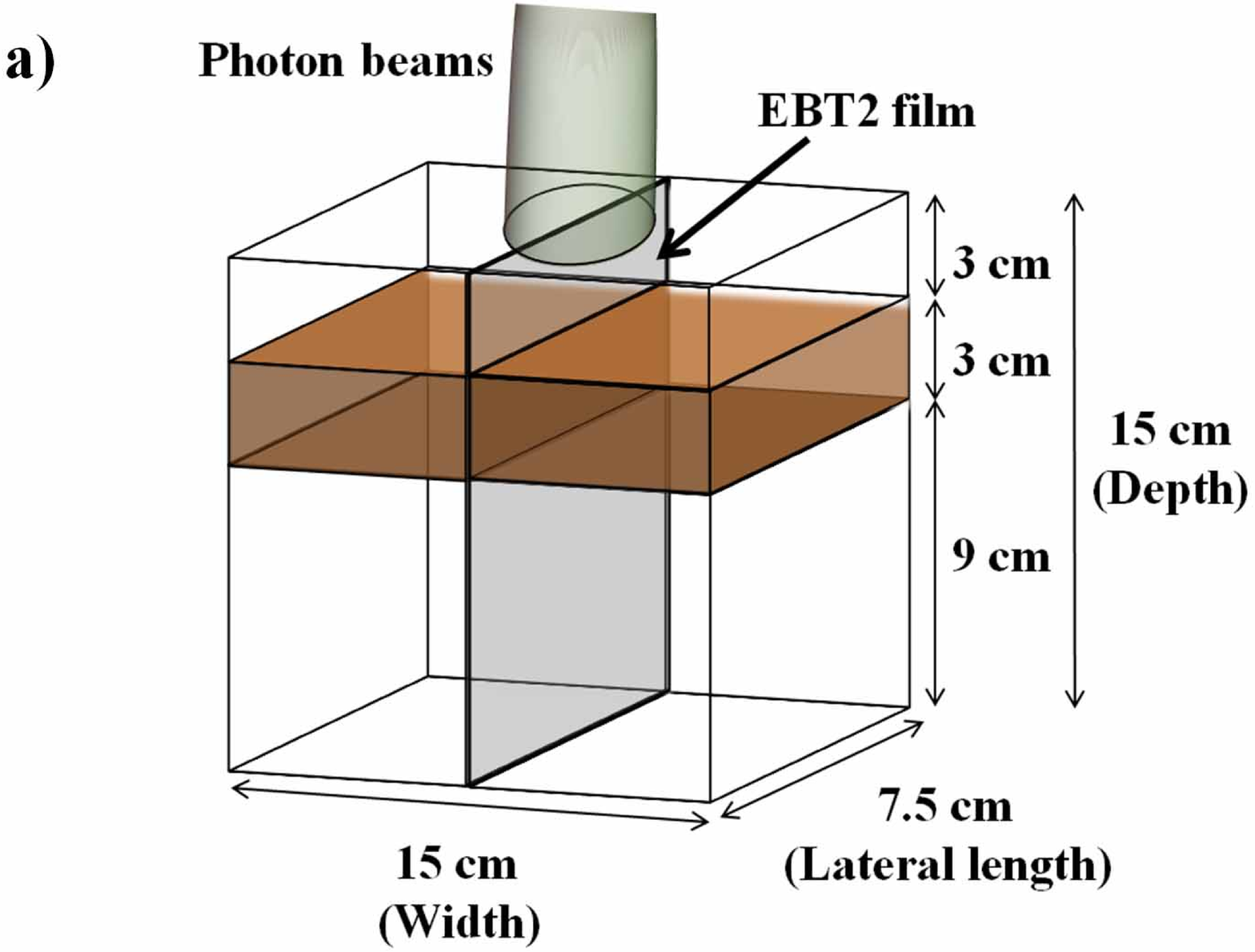, width=3.1in}
\epsfig{file=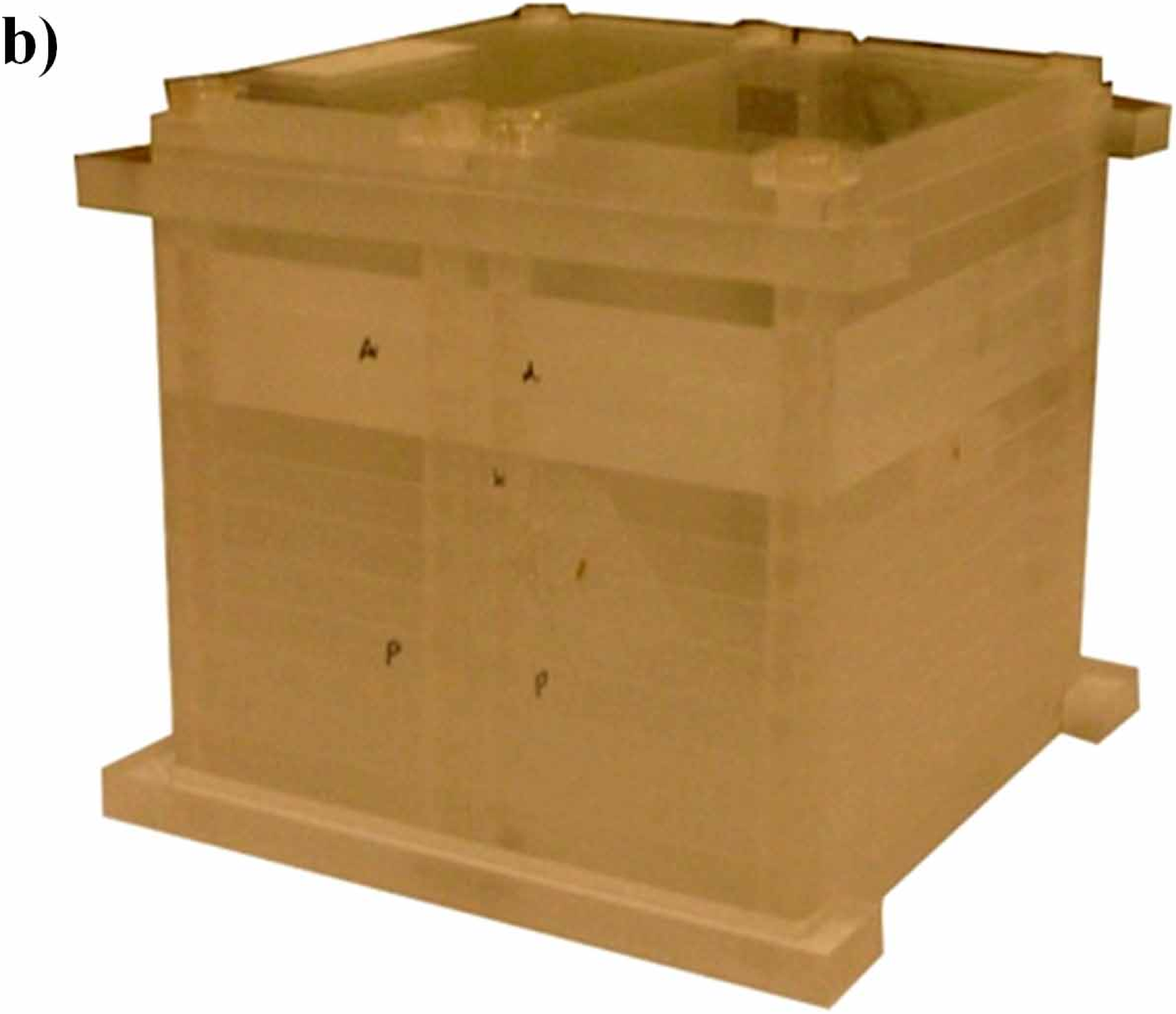, width=3.in}
\caption{(Color online) The inhomogeneous phantom has a dark horizontal part 
which represents either the Balsa wood or bone equivalent materials. 
Photon beams are irradiated from the top in parallel to the film which is 
inserted vertically in the middle of the phantom. 
A photo of the phantom is shown in (b).} 
\label{fig1}
\end{figure}

Figure \ref{fig1} shows a schematic diagram 
and a photo of the phantom used in this work. 
In Fig. \ref{fig1} (a), the horizontal slab of 3 cm thickness 
inserted in the phantom denotes 
the part for lung or bone equivalent materials. 
Gafchromic EBT2 film is inserted vertically 
in the middle of the phantom. 
The phantom is composed of polystyrene material 
except for the inhomogeneous region. 
The size of the phantom is 15 cm in width 
and 7.5 cm in lateral length. 
The total depth of the phantom is 15 cm, 
which consists of 3 cm of polystyrene, 
3 cm of Balsa wood or soft bone equivalent material and 9 cm of polystyrene. 
Polystyrene is used for the soft tissue. 
Balsa wood and soft bone equivalent materials 
are used to simulate low-density medium and high-density medium, respectively. 
The cross section of Balsa wood has many microscopic holes 
and is similar to the lung. 
We have used the Balsa wood with density of 0.116 g/cm$^{3}$. 
This value of density is about 50\% 
lower than the one used in Ref. \cite{r_22}. 
This choice is good to observe the difference 
between the DTP and the DTF. 
For high density materials, we have used soft bone equivalent materials. 
According to White {\em et al}. \cite{r_32} 
the average density of the spine is 1.320 g/cm$^{3}$. 
We made a soft bone equivalent material comprised of epoxy resin (40\%), hardener (20\%) 
and calcium carbonate (40\%). 
Compounds of epoxy resin and calcium carbonate can be appropriate materials to substitute 
for spinal tissues because their compositions are similar to each other. 
Such soft bone equivalent materials used in our study have a density of 1.363 g/cm$^{3}$.

%%%%%%%%%%%%%%%%%%%%%%%%%%%%%%%%%%%%%%%%%%%%%%%%%
%%%%%%%%%%%%%%%%%%%%%%%%%%%%%%%%%%%%%%%%%%%%%%%%%%%

The EBT2 film was wedged tightly in the middle of the phantom to minimize dose perturbation. 
Then the phantom with the EBT2 film was irradiated by photon beams with 5, 10 and 30 mm cone sizes. 
The source to surface distance (SSD) was chosen as 80 cm, and irradiation dose was set to 400 cGy 
at the maximum dose position. 
Gafchromic EBT2 films were calibrated and scanned after 24 hours 
by the scanner Epson V700 PHOTO (Epson USA). 
The optical density was analyzed with OmniPro-IMRT. 
The CyberKnife treatment system Multiplan v2.0.4 was 
also used to calculate the PDD for comparison 
with the results from GEANT4 and measurements.

\subsection{GEANT4 simulation}

GEANT4 is a simulation tool kit written in C++ language, 
which enables the simulation of propagation of particles that interact with the materials 
and/or other particles. 
It is widely used in many different scientific fields, 
such as high energy and nuclear physics \cite{r_33}, 
environment radiation detections \cite{r_34}, 
medical physics \cite{r_29, r_31, r_35}, 
and other applications \cite{r_36, r_37}.

In this work, the simulations of the electromagnetic (EM) processes 
in GEANT4 have been performed with 
Low Energy package \cite{r_38}. 
This Low Energy package includes the photo-electric effect, Compton scattering, 
Rayleigh scattering, gamma conversion, bremsstrahlung and ionization. 
Fluorescence of excited atoms is also considered. 
The current implementation 
of low energy processes is valid for energies down to 250 eV. 
Data used for the determination of cross sections 
and for sampling the final state are extracted 
from a set of publicly distributed evaluated data libraries, 
EADL (Evaluated Atomic Data Library) \cite{r_39}, 
EEDL (Evaluated Electrons Data Library) \cite{r_40}, 
EPDL97 (Evaluated Photons Data Library) \cite{r_41}, 
and stopping power data \cite{r_42, r_43, r_44, r_45}.

%%%%%%%%%%%%%%%%%%%%%%%%%%%%%%%%%%%%%%%%%%%%%%%%%
\begin{figure}[tbp]
\epsfig{file=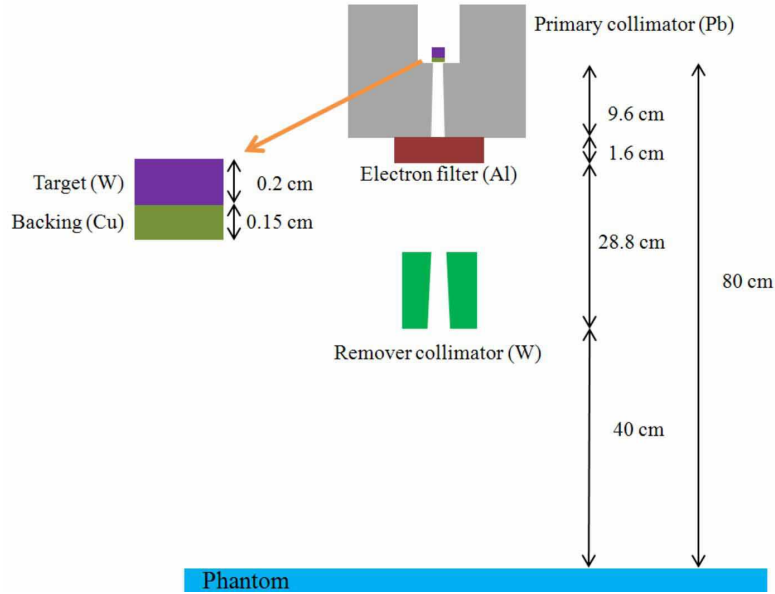, width=5in}
\caption{(Color online) A schematic diagram of 
the CyberKnife system used in the simulations.} 
\label{fig2}
\end{figure}
%%%%%%%%%%%%%%%%%%%%%%%%%%%%%%%%%%%%%%%%%%%%%%%%%%%

In our previous study \cite{r_31}, 
we modeled a CyberKnife system at the Seoul Uridul Hospital 
and simulated the photon beams produced by 
6 MV CyberKnife using GEANT4 (version 9.2.p01). 
Figure \ref{fig2} shows a schematic diagram of 
the CyberKnife system used in the simulations. 
To make the simulation more efficient, 
the calculation procedure was split into two steps. 
The CyberKnife produces electrons impinging on the tungsten target, 
which creates bremssthlung photons. 
In the first step, we scored the energy and 
momentum distribution of the photons for 5 to 60 mm cone sizes 
arriving at the surface of the water phantom of SSD of 80 cm. 
As the cone sizes vary, 
different energy and momentum distributions are obtained. 
In the second step, the photon beams were generated at the surface of the water phantom 
in the direction towards the phantom material by using the energy and 
angular distributions of the photons as scored in the first step. 
This two step method can save computation time greatly. PDD distributions, output factors and off axis ratios 
were also calculated in the second step. 
In Ref. \cite{r_31}, we found that our calculated values of 
these quantities in this two step method agreed with 
our measured values with errors less than 2\%.

In this work, we used the same photon beam distribution as scored in Ref. \cite{r_31}. 
We modeled inhomogeneous phantoms with and without the EBT2 film. 
An EBT2 film consists of 5 layers with different thickness and element compositions. 
Detailed information of the film used in our simulations is available 
from the product specification \cite{r_46}. 
For the simulations with the EBT2 film, 
the size of the scoring voxel is chosen to have the thickness of 
the active layer of the EBT2 film which is 30 $\mu$m. 
The lateral lengths of the voxels are set to 0.5, 1 and 3 mm 
for 5, 10 and 30 mm of cone sizes, respectively. 
The same conditions for the positions and sizes of the scoring voxels 
are used for the case of the simulations 
without the EBT2 film. First, we have performed 
benchmark simulations to confirm the consistency 
between our GEANT4 simulations and some previous studies \cite{r_12, r_22} done by others. 
Both homogeneous \cite{r_12} and inhomogeneous \cite{r_22} phantoms were considered. 
In this benchmark simulation also, 
we have considered phantoms with and without the EBT film included. 
In the simulations the production cut for the photons 
and the electrons was set to 10 keV and 100 keV, respectively. 
Each simulation was repeated ten times with different random number seeds. 
The results from these ten different simulations were averaged 
to get the final results and statistical errors. 
Simulation results are compared with our experimental results.

\section{RESULTS AND DISCUSSION}

\subsection{Benchmark of our GEANT4 simulations with previous studies}

In order to check the validity of our GEANT4 simulations, 
we have first calculated PDDs 
for homogeneous and inhomogeneous phantoms 
with conditions as given in Refs. \cite{r_12} and \cite{r_22}. 
As mentioned earlier, simulations have been performed in two cases; 
with and without the EBT film. 
First, the simulations are done with 
only polystyrene phantom without the EBT film. 
In this case, the absorbed doses were calculated 
in the polystyrene phantom. 
As noted in the Introduction, 
the calculated doses in this way are referred as DTP. 
Second, the simulations are done by taking into account 
the presence of the EBT film. 
The doses calculated in this way are referred to as DTF.

%%%%%%%%%%%%%%%%%%%%%%%%%%%%%%%%%%%%%%%%%%%%%%%%%
\begin{figure}[tbp]
\epsfig{file=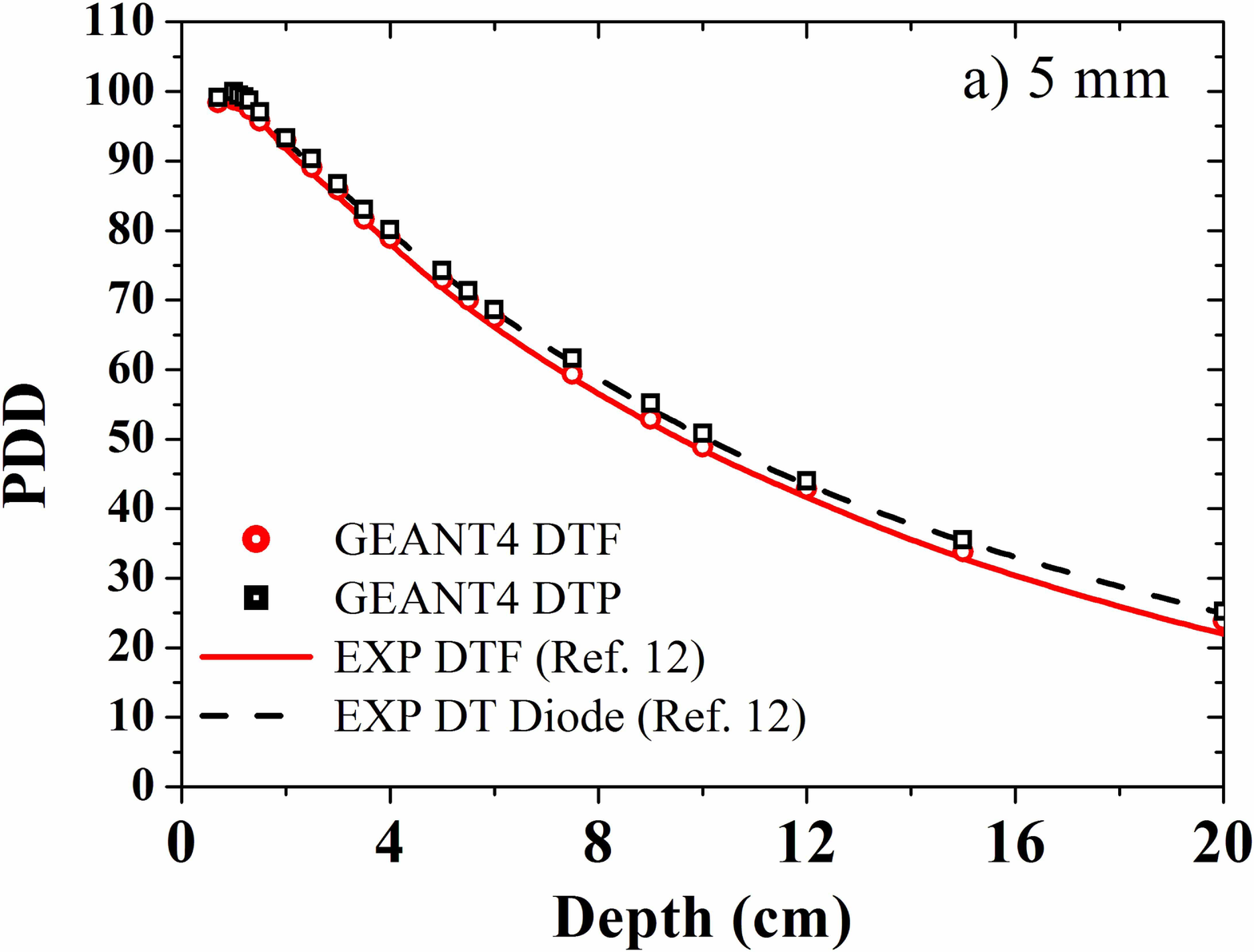, width=3.1in}
\epsfig{file=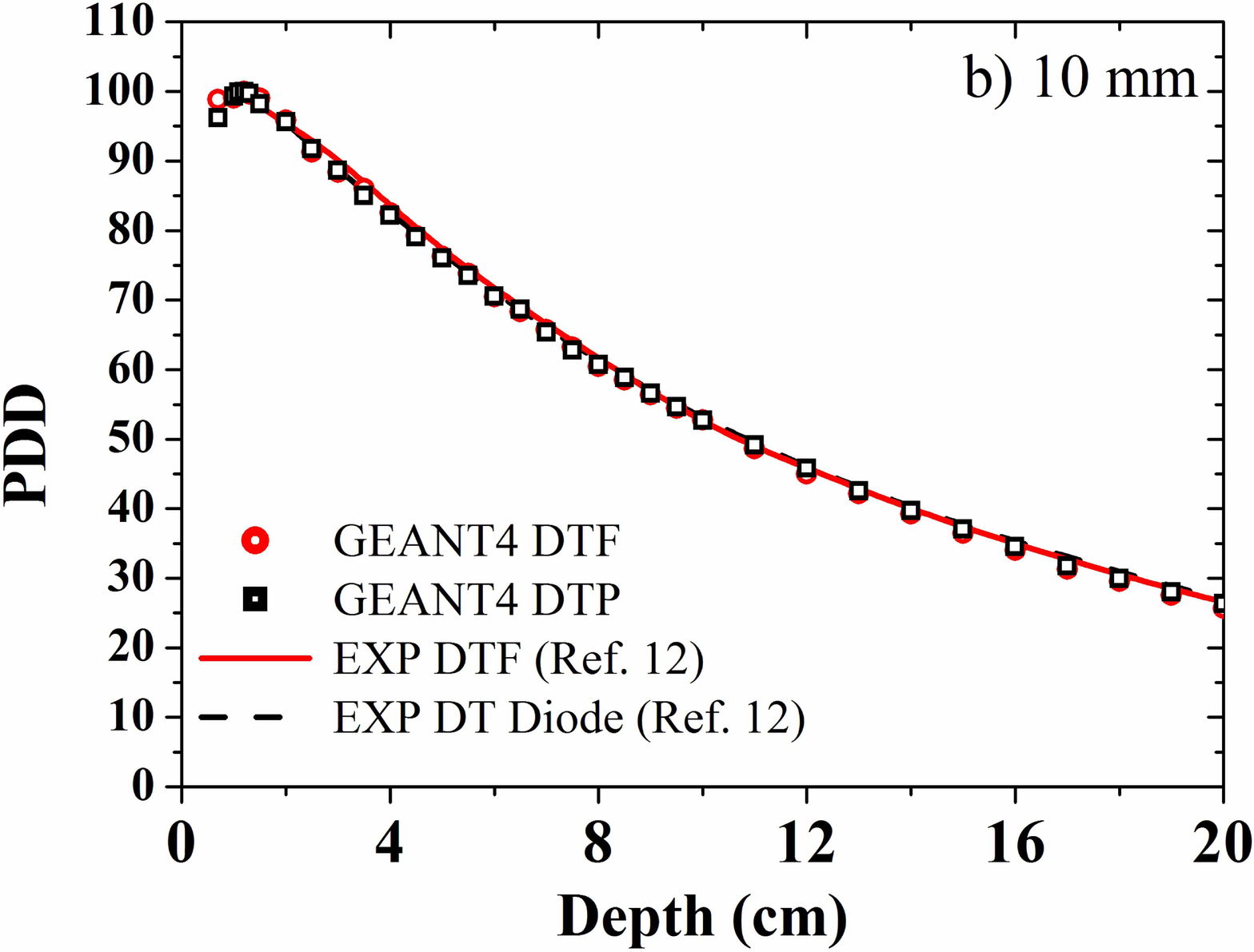, width=3.1in}
\caption{(Color online) PDD distributions in water phantom for 5 and 10 mm cone sizes. 
The solid and the dashed lines represent the measured dose \cite{r_12} values 
with film and diode detectors, respectively. 
The open circles and the open squares denote the calculated doses 
with and without the EBT film, respectively. DTF (DTP) refers to dose to film (dose to phantom). 
For a cone size of 10 mm the DTP and the DTF are the same, but for a small cone size 
of 5 mm the DTF is smaller than the DTP by about 5\%. 
For both cases, our GEANT4 calculations agree with the measured values.} 
\label{fig3}
\end{figure}
%%%%%%%%%%%%%%%%%%%%%%%%%%%%%%%%%%%%%%%%%%%%%%%%%%%

EBT films have different types and lot numbers. 
It was reported that the calculations of the absorbed dose have 
dependency on the type of EBT films 
if the energy of the photons is below 0.3 MeV \cite{r_47}. 
To check the dependency of PDD calculations on the type of EBT films 
for the case of CyberKnife we used, 
we have considered three different types of EBT films: 
EBT1, EBT2 (lot020609) and EBT2 (lot031109), 
but discrepancies are not observed within 
the calculation errors of 0.5\% for all the cases considered in this work. 
It may be because the photon beam energy from 
the 6MV CyberKnife is much higher than the photon beams used in Ref. \cite{r_47}. 
Therefore, we shall not distinguish different types of the EBT films in this work 
and refer to the EBT2 film used in this work simply as the EBT film.

Let us first compare the PDD obtained from our simulations 
with the published PDD \cite{r_12} in homogeneous water phantom. 
Figure \ref{fig3} (b) shows good agreements between our calculations 
and the measurements obtained for 10 mm cone size regardless of the presence of the EBT film. 
The difference between the doses obtained from GEANT4 and 
the measured doses is smaller than 1.5\% on the average for 10 mm cone size. 
Similar good agreements are observed for 30 mm cone size, 
though the comparison is not shown in Fig. \ref{fig3}. 
For 5 mm cone size, the doses obtained from GEANT4 with EBT film (without EBT film) 
still agree well with the measured doses with EBT film (with diode detector) 
with errors of 3.6\% (1.7\%), respectively. 
However, the PDD with EBT film is slightly smaller than that without the EBT film. 
The average ratio of the measured dose to diodes (DT diodes) to the measured DTF was 1.055. 
A similar value of 1.035 is obtained for the average ratio of the GEANT4 DTP to the GEANT4 DTF.

%%%%%%%%%%%%%%%%%%%%%%%%%%%%%%%%%%%%%%%%%%%%%%%%%
\begin{figure}[tbp]
\epsfig{file=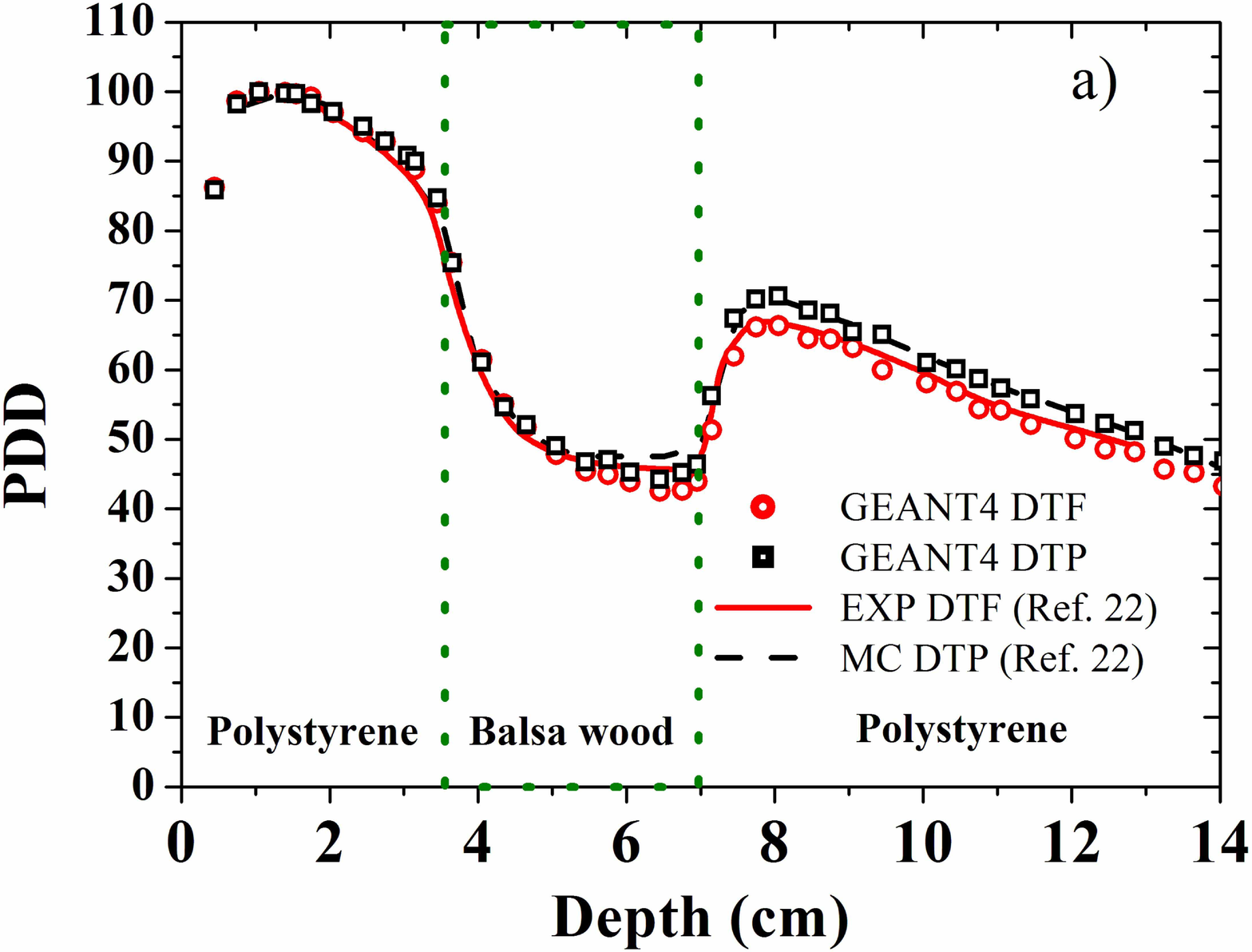, width=3.1in}
\epsfig{file=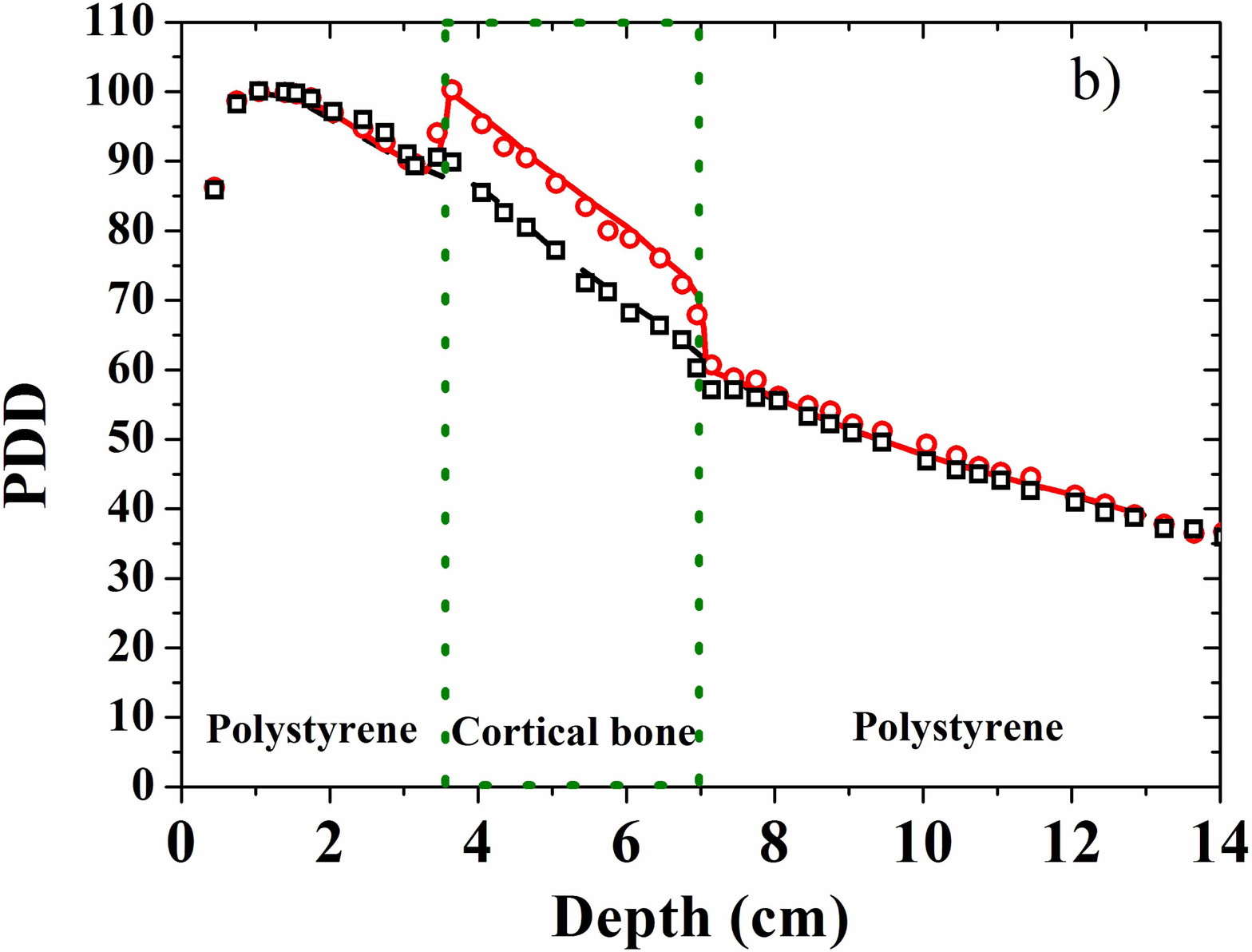, width=3.1in}
\caption{(Color online) PDD distributions in an inhomogeneous phantom 
for a beam of 10 mm cone size. 
The solid and the dashed lines represent dose measured with EBT film 
and that calculated without the EBT film, respectively \cite{r_22}. 
The open circles and the open squares denote our calculated doses 
with and without EBT film, respectively.} 
\label{fig4}
\end{figure}
%%%%%%%%%%%%%%%%%%%%%%%%%%%%%%%%%%%%%%%%%%%%%%%%%%%

Second, we compare our calculated PDD with the calculated 
and measured PDD \cite{r_22} for inhomogeneous phantoms. 
Figure \ref{fig4} (a) and (b) show the PDD in inhomogeneous phantom 
with the Balsa wood and cortical bone, respectively. 
In the case when the Balsa wood is inserted in the phantom, 
our calculations reproduce well the PDD measured 
and calculated in Ref. \cite{r_22}. 
Figure \ref{fig4} (a) shows some difference between the doses with and without EBT films 
in our calculations as observed in the previous results \cite{r_22} 
in the region beyond the Balsa wood. 
Also, Fig. \ref{fig4} (b) shows that when cortical bone equivalent material is inserted 
in the phantom our calculations with and without EBT film reproduce 
the PDD measured with EBT film \cite{r_22} 
and the PDD calculated without EBT film, respectively. 
Thus previous calculations and measurements \cite{r_22} 
are well reproduced by our calculations, 
which confirms the accuracy of our GEANT4 calculations 
for both cases with and without the EBT film. 
This results also shows that the presence of the EBT film affects 
the value of PDD for a small cone size of 5 mm 
and needs to be properly taken into consideration in the simulation.

\subsection{Comparison of simulations with experiments with inhomogeneity of water-lung-water}

%%%%%%%%%%%%%%%%%%%%%%%%%%%%%%%%%%%%%%%%%%%%%%%%%
\begin{figure}[tbp]
\epsfig{file=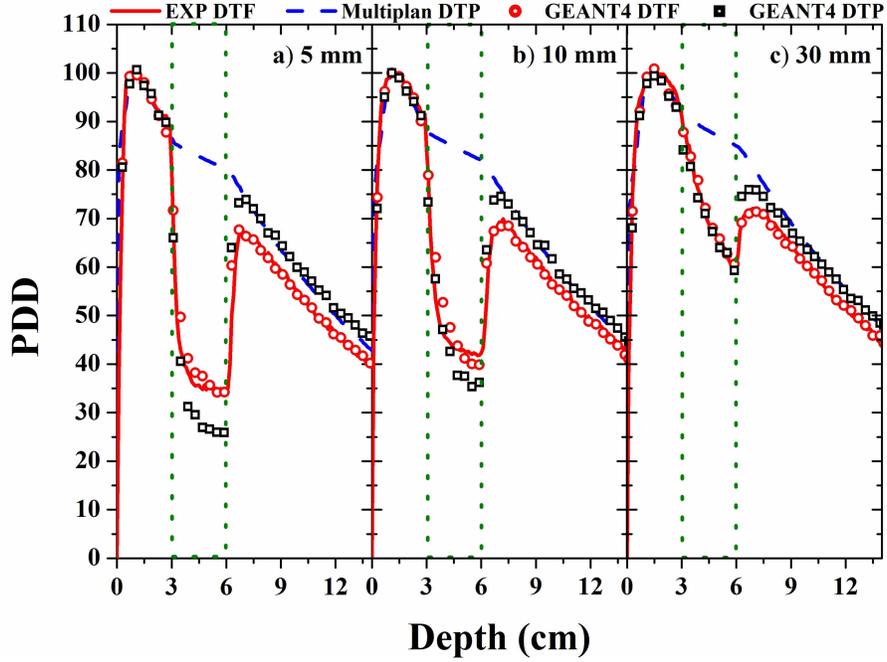, width=5in}
\caption{(Color online) PDD distributions in an inhomogeneous phantom 
with Balsa wood for three field sizes (5, 10 and 30 mm). 
The solid line denotes the PDD measured with the EBT film. 
The dashed line represents the calculated dose values by using Multiplan. 
The open circles and the open squares denote the calculated PDD 
with and without the EBT film, respectively. 
GEANT4 DTF reproduces well EXP DTF.}
\label{fig5}
\end{figure}
%%%%%%%%%%%%%%%%%%%%%%%%%%%%%%%%%%%%%%%%%%%%%%%%%%%

%%%%%%%%%%%%%%%%%%%%%%%%%%%%%%%%%%%%%%%%%%%%%%%%%
\begin{figure}[tbp]
\epsfig{file=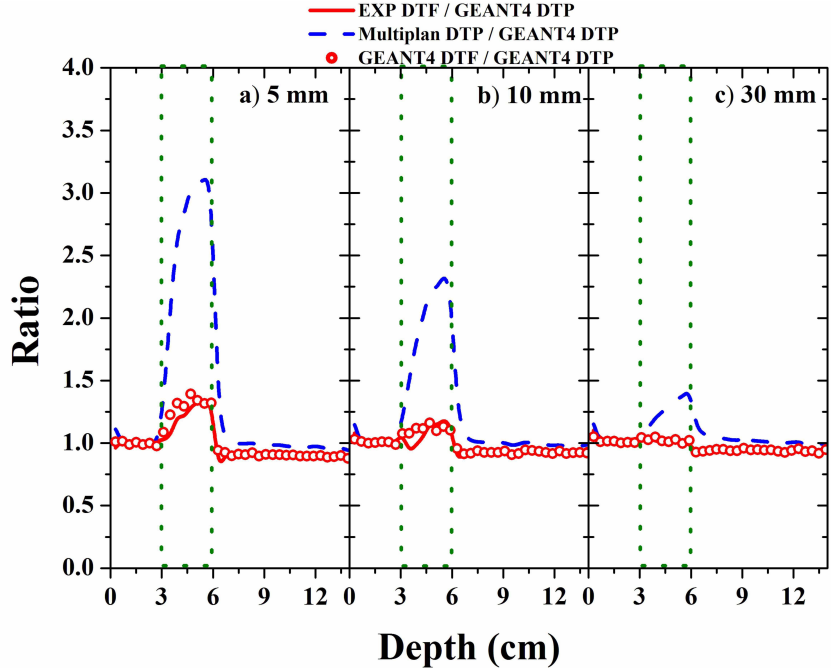, width=5in}
\caption{(Color online) The ratios of the three different PDD's to the GEANT4 DTP 
are plotted as a function of the depth. 
The ratio of the measured DTF to GEANT4 DTP is shown by the solid line 
which agrees with that of the GEANT4 DTF to the GEANT4 DTP denoted by the open circles. 
The ratio of the DTP calculated from Multiplan to 
the calculated DTP is plotted by the dashed line.} 
\label{fig6}
\end{figure}
%%%%%%%%%%%%%%%%%%%%%%%%%%%%%%%%%%%%%%%%%%%%%%%%%%%

We performed the calculations and measurements of PDDs for 
an inhomogeneous phantom with Balsa wood for three field sizes of 5, 10, 30 mm, 
and the results are represented in Fig. \ref{fig5}. 
The PDD calculated with the EBT film (GEANT4 DTF) reproduce well the PDD measured 
with the EBT film (EXP DTF). 
The average percentage errors between the GEANT4 DTF and the EXP DTF 
for 5, 10 and 30 mm of cone sizes are 2.20\%, 0.43\% and 0.08\%, respectively. 
However, Multiplan cannot reproduce the measured dose values 
in both the Balsa wood region and beyond the Balsa wood region. 
Large discrepancies are observed especially in the Balsa wood region. 
The average ratio of the PDD obtained from Multiplan to those measured with EBT film 
for 5, 10 and 30 mm of field sizes are 2.1, 1.8 and 1.2, respectively, 
which shows larger discrepancies for smaller cone sizes. 
The same tendency can also be seen in Ref. \cite{r_22}. 
Figure \ref{fig5} also shows there is 
a significant difference between the PDD calculated 
without the EBT film (GEANT4 DTP) 
and the PDD measured with the EBT film. 
This clearly shows that the proper inclusion 
of the EBT film in the simulation 
is necessary for accurate calculations of the dose. 
In Fig. \ref{fig6} we have compared the GEANT4 DTP 
with the calculated and measured DTF. 
In the Balsa wood region, the average ratios of the EXP DTF 
to the GEANT4 DTP are 1.29, 1.11, 
and 1.02 for 5, 10 and 30 mm cone sizes, respectively. 
The average ratios of the DTP obtained from 
Multiplan to the GEANT4 DTP are 2.66, 1.95, 
and 1.26 for 5, 10, and 30 mm cone sizes, respectively. 
In the region beyond Balsa wood, the average ratios of the EXP DTF 
to the GEANT4 DTP are 0.90, 0.93, 
and 0.94 for 5, 10, and 30 mm cone sizes, respectively, 
whereas the average ratios of the DTP from Multiplan 
to the GEANT4 DTP are 0.99, 1.01, and 1.02, respectively, for 5, 10, and 30 mm cone sizes. 
When the EBT film is included, the doses 
in the Balsa wood region are enhanced and those 
in the region beyond Balsa wood are reduced. 
As the cone size increases, the discrepancy decreases, 
which shows that accurate estimates and measurements of doses 
are important particularly for small field sizes. 
Similar features are observed in Refs. \cite{r_18}, \cite{r_21} and \cite{r_22}.

%%%%%%%%%%%%%%%%%%%%%%%%%%%%%%%%%%%%%%%%%%%%%%%%%
\begin{figure}[tbp]
\epsfig{file=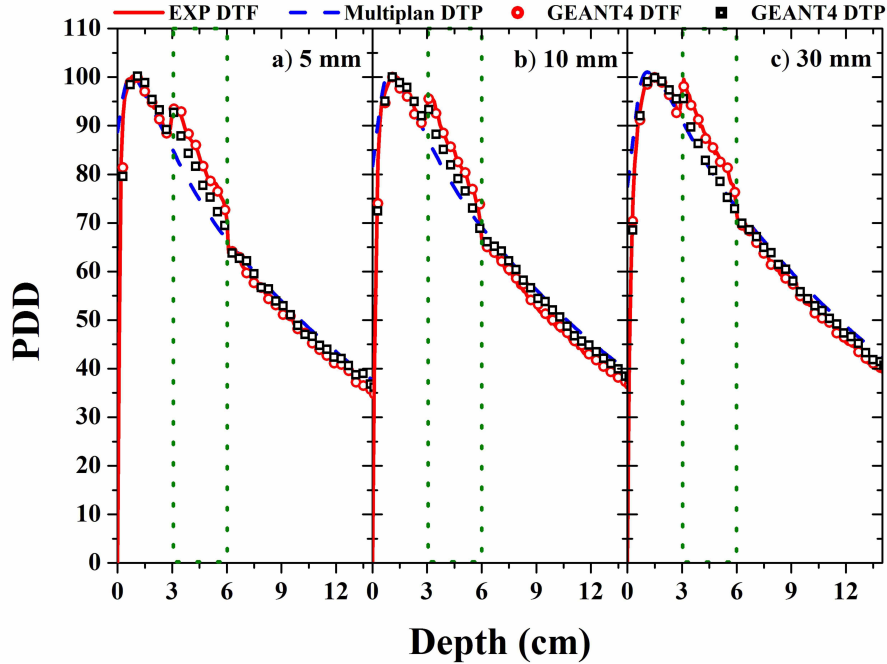, width=5in}
\caption{(Color online) PDD distributions for the inhomogeneous phantom 
with soft bone material for three field sizes (5, 10 and 30 mm). 
The solid line denotes the experimental PDD measured with EBT film. 
The dashed line represents the PDD calculated with Multiplan. 
The open circles and open squares denote the doses calculated by GEANT4 
with and without the EBT film, respectively. The open circles (GEANT4 DTF) 
and the experimental DTF agree with each other very well.}
\label{fig7}
\end{figure}
%%%%%%%%%%%%%%%%%%%%%%%%%%%%%%%%%%%%%%%%%%%%%%%%%%%

\subsection{Comparison of simulations with experiments with inhomogeneity of water-bone-water}

\begin{figure}[tbp]
\epsfig{file=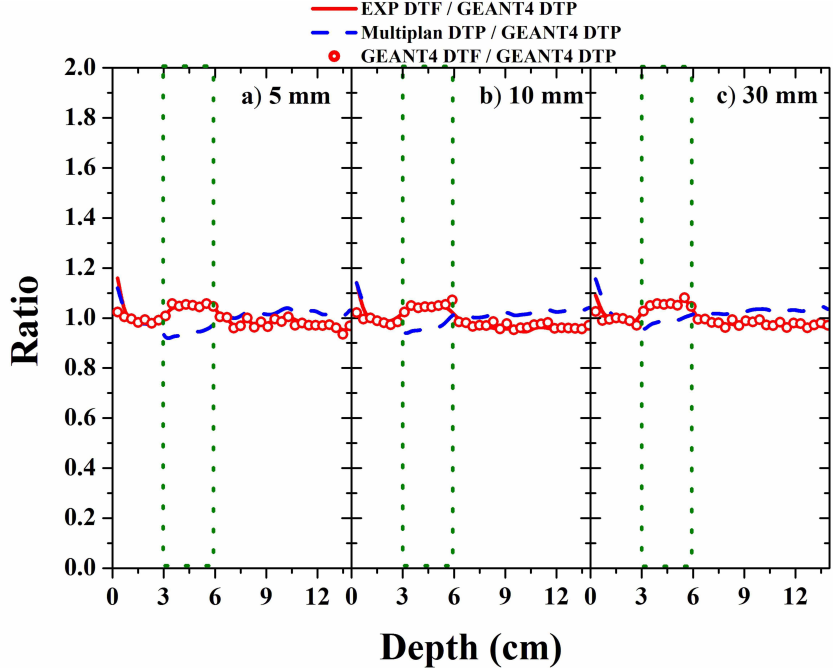, width=5in}
\caption{(Color online) The ratios of the GEANT4 DTF 
and measured EXP DTF to the GEANT4 DTP are represented. 
The ratios of the calculated DTP obtained 
from Multiplan to the GEANT4 DTP are also shown.} 
\label{fig8}
\end{figure}

The PDDs for an inhomogeneous phantom 
with the soft bone are presented in Fig. \ref{fig7}. 
The average percentage errors between 
the GEANT4 DTF and the measured EXP DTF 
for 5, 10, 30 mm of cone sizes are 0.41\%, 0.30\% 
and 0.09\%, respectively. 
On the other hand, Multiplan cannot reproduce 
the dose values in the soft bone regions. 
In contrast to the inhomogeneous phantom with Balsa wood, 
the discrepancies appear only in the soft bone region. 
Figure \ref{fig8} shows the differences between the dose with the EBT film 
and that without the EBT film. 
In the soft bone region, the average ratios of 
the GEANT4 DTF (and experimental DTF) to the GEANT4 DTP 
are 1.05, 1.05 and 1.05 for 5, 10 and 30 mm cone sizes, respectively. 
The average ratios of the DTP obtained from Multiplan to the GEANT4 DTP are 0.94, 0.96, 
and 0.99 for 5, 10, and 30 mm cone sizes, respectively. 
It can be seen that the ratios of the DTF to DTP are nearly constant as 1.05 
and independent of the cone sizes.

%%%%%%%%%%%%%%%%%%%%%%%%%%%%%%%%%%%%%%%%%%%%%%%%%
%%%%%%%%%%%%%%%%%%%%%%%%%%%%%%%%%%%%%%%%%%%%%%%%%%%

Difference between the doses to water and 
the doses to other media was studied by Siebers {\em et al}. \cite{r_23}. 
They suggested correction factors for various beam energies and materials. 
For the case of the 6MV photon beams, 
correction factors needed for the lung, 
ICRU tissue, soft bone and cortical bone are 1, 1.101, 1.035 and 1.117, respectively. 
Wilcox {\em et al}. \cite{r_22} showed that the values of 
the DTP multiplied by the correction factor 1.117 
in the cortical bone region agree with those of 
the doses obtained with the EBT film. 
From our study, the average ratios of DTF to DTP 
are obtained to be 1.049 $\pm$ 0.015 and 1.129 $\pm$ 0.015 
for soft bone and cortical bone, respectively. 
Our ratios (1.049 and 1.129) and 
the correction factors (1.035 and 1.117) of Ref. \cite{r_23} 
agree with each other within 1.3\%.

\section{Conclusion}

We have studied the PDD distributions in an inhomogeneous phantom 
with low and high density materials. 
Measurements were performed with Gafchromic EBT film 
using the 6MV CyberKnife system 
at the Seoul Uridul Hospital. 
The CyberKnife system was modeled 
and PDDs were calculated with GEANT4 code. 
Simulations have been performed with and without the EBT film. 
We find that it is important to include 
the presence of EBT film to accurately calculate the PDD values, 
in particular, for small field sizes. 
Comparison of the simulation results 
with the measured ones with EBT film shows that our calculations 
agree with the measurements within an error of 2.2\% 
for all the cases considered here. 
However, Multiplan provided by the CyberKnife cannot reproduce the measured PDD. 
We confirm the correction factors suggested by Ref. \cite{r_23} 
for the case when high density materials 
(soft bone or cortical bone) are inserted in the phantoms. 
The doses to phantom (DTP) without the EBT film can be different 
from the doses to film (DTF) by 29\% for 5 mm cone size. 
Thus the presence of the EBT film needs to be 
taken into account in particular for small fields.

\begin{acknowledgments}
This work was supported by the National Research Foundation (NRF) of Korea 
grant funded by the Korean government (MEST) (No. 2012000486) 
and WCU program through NRF funded by the MEST (R31-2008-10029).
\end{acknowledgments}

%\begin{references}

%\end{references}

%\section{Table}

%\section{Figures}

\end{document}